\documentclass[reqno,10pt,centertags]{article}
\usepackage{amsmath,amsthm,amscd,amssymb,latexsym,upref}
\usepackage{amsxtra,amssymb,amsthm,amsmath,latexsym}

\newcommand{\lb}{\label}

\newcommand{\all}{\forall}

\newcommand{\Int}{\int\limits}

\renewcommand{\l}{\lambda}
\renewcommand{\d}{\partial}

\renewcommand{\b}{\beta}

\allowdisplaybreaks
\numberwithin{equation}{section}

\newtheorem{theorem}{Theorem}[section]
 
\newtheorem{lemma}{Lemma}[section] 
 
\theoremstyle{definition}

\theoremstyle{remark}
\newtheorem{remark}[theorem]{Remark}

\begin{document}

Applic. Analysis, 81, N4, (2002), 929-937.

\title
{An inverse problem for the heat equation II 
\thanks{Primary: 35R30, 81U40; Secondary: 47A40}
\thanks{inverse problem, conductivity, flux, property C, potential, Sturm-
Liouville equations}}
\author{A.~G.~Ramm\\ 
LMA/CNRS, 31 Chemin J.Aiguier\\
Marseille 13402, France\\
and Department of Mathematics,\\
Kansas State University     \\
Manhattan, Kansas,
66506-2602, USA\\
email:  ramm@math.ksu.edu\\
}
\date{} 
\maketitle
\begin{abstract} Completeness of the set of products of the derivatives of the 
solutions to the equation $(av')'-\l v=0,\ v(0,\l)=0$ is proved. This property 
is used to prove the uniqueness of the solution to an inverse problem of finding 
conductivity in the heat equation $\dot{u}=(a(x)u')',\ u(x,0)=0,\ u(0,t)=0,\ 
u(1,t)=f(t)$ known for all 
$t>0$, from the heat flux $a(1)u'(1,t)=g(t)$. Uniqueness of the solution to this 
problem is proved. The proof is based on Property C. It is proved 
the inverse that the inverse
problem with the extra data (the flux) measured at the point, where
the temperature is kept at zero, (point $x=0$ in our case) does
not have a unique solution, in general. 
\end{abstract}

\maketitle

\section{Introduction}\lb{s1}
In a study of one and multidimensional inverse problems Property C 
(completeness of the set of products of solutions to some
homogeneous 
equations) was introduced and has been used extensively \cite{R3},\cite{R5}. 
Completeness of the set of products of the eigenfunctions of some Sturm-
Liouville 
equations was studied in \cite{Lev1}, and, in a more general context, in 
\cite{R3}-\cite{R6}. In this paper we prove completeness of the set 
of products of the derivatives of solutions to a second-order equation and use 
it to study an inverse problem for it. A similar idea was used in \cite{R5} in 
the multidimensional case. It is of interest to see that property C
allows one to decide which measurements are most informative: we prove,
using this property, that the measurements of the heat flux at the
point $x=1$ allow one to uniquely determine the conductivity, while
 the measurements of the heat flux at the point  $x=0$ do not
allow one to determine the conductivity  uniquely, in general. 

Denote $\dot{u}:=\frac{\d u}{\d t}$, $u':=\frac{\d u}{\d x}$. Consider the 
following problem:
\begin{equation}\lb{ip1}
\dot{u}=(a(x)u')',\ 0\leq x\leq 1,\ t>0,
\end{equation}
\begin{equation}\lb{ip2}
u(x,0)=0,
\end{equation}
\begin{equation}\lb{ip3}
u(0,t)=0,\ u(1,t)=f(t),
\end{equation}
\begin{equation}\lb{ip4}
a(1)u'(1,t)=g(t).
\end{equation}
Assume:
$$
f\in L^1([0,1]),\ f\not\equiv0,\ a\in W^{2,1}([0,1]), a>0, 
$$ 
where $W^{2,1}$ is the Sobolev space of functions twice differentiable in 
$L^1([0,1])$. The function $u(x,t)$ is the temperature of a 
medium, $a(x)$ is its conductivity, $g(t)$ is the heat flux (through the 
point $x=1$), which can be measured experimentally. The inverse problem is 
to find the  conductivity of the medium from the flux measurements. Formulas for 
finding conductivity at the boundary in the multidimensional case were derived 
in \cite{R6}.

Applying the Laplace transform to (\ref{ip1})-(\ref{ip3}) one gets the equation 
$(a(x)v')'-\l v=0$ ($\l$ is the spectral parameter) with certain boundary 
conditions and extra data. Our uniqueness result for (\ref{ip1})-(\ref{ip4}) 
follows from the completeness of the set of products of the derivatives of the 
solutions to this equation. It is interesting to note that if (\ref{ip4}) is 
replaced by  
\begin{equation}\lb{ip5}
a(0)u'(0,t)=h(t),
\end{equation}
the result is no longer valid and, in fact, only 'one half' of $a(x)$ can be 
recovered, in general. Similar non-uniqueness result was obtained in \cite{R7}. 

The structure of this paper is the following. In section \ref{s2} we reduce  
the completeness of the set of products of the derivatives of the solutions to 
Property C for a Sturm-Liouville equation with the third boundary condition. 
Though Property C for this equation follows from a result in \cite{Lev1}, we 
give a simple alternative proof. In section \ref{s3} we use Property C to prove 
the uniqueness result for (\ref{ip1})-(\ref{ip4}) and also a non-uniqueness for 
(\ref{ip1})-(\ref{ip3}),(\ref{ip5}).

\section{Completeness of the set of products of the derivatives of the 
solutions}\lb{s2}

\begin{theorem}\lb{propC} Let $v_i$, $i=1,2$, solve the equations
\begin{equation}\lb{eqvi}
(a_i(x)v'_i)'-\l v_i=0,\ 0\leq x\leq1,\ v_i(0,\l)=0,
\end{equation}
where $0<a_i\in W^{1,2}([0,1])$. Then the set of products $\{v'_1(\cdot,\l) 
v'_2(\cdot,\l)|\all\l>0\}$ is complete in $L^1([0,1])$.
\end{theorem}
To prove this theorem we need some lemmas.
\begin{lemma}\lb{der} A function $w$ solves the problem
\begin{equation}\lb{eqw}
w''-\l a^{-1}(x)w=0,\ w'(0,\l)=0
\end{equation}
if and only if $v(x):=\Int_0^x a^{-1}(t)w(t)dt$ solves the problem
\begin{equation}\lb{eqv}
(av')'-\l v=0,\ v(0,\l)=0.
\end{equation}
\end{lemma}
\begin{proof}Let $v$ satisfy (\ref{eqv}) and let $w:=av'$. Then 
$v(x):=\Int_0^x a^{-1}(t)w(t)dt$, and a differentiation of
(\ref{eqv}) yields 
(\ref{eqw}).

Conversely, let $w$ solve (\ref{eqw}), and define $v(x):=\Int_0^x 
a^{-1}(t)w(t)dt$. Integrating (\ref{eqw}) from $0$ to $x$ one gets $(av')'(x)-\l 
v(x)=0$ and $v(0)=0$. 
\end{proof}
By Lemma \ref{der} the set of the derivatives of the solutions to (\ref{eqv}) 
coincides, up to a factor $a^{-1}(x),$ 
 with the set of the solutions to (\ref{eqw}). Our next step is to 
reduce (\ref{eqw}) to a Sturm-Liouville equation.
\begin{lemma}\lb{iso} Under the substitution $\xi(x):=\Int_0^xa^{-1/2}(t)dt$, 
$w(x)=a^{1/4}(x)z(\xi(x))$ the problem (\ref{eqw}) is equivalent to the 
problem:
\begin{equation}\lb{eqz}
\ddot{z}-q(\xi)z-\l z=0,\ \dot{z}(0)+hz(0)=0, 
\end{equation}
where $\dot{z}:=\frac{dz}{d\xi}$, $q(\xi(x)):=\frac3{16}a^{-1/2}(x)(a')^2(x)-
\frac14a''(x)$, $h:=\frac14a^{-1/2}(0)a'(0)$.
\end{lemma}
\begin{proof} Make the following substitutions
$$
\xi(x):=\Int_0^x\phi(t)dt,\ w(x):=z(\xi(x))\psi(x),
$$ 
where the functions $\phi$ and $\psi$ are to be determined.
Let us derive an equation for $z$ assuming that $w$ satisfies 
(\ref{eqw}). One has 
$$
w'=\psi'z+\psi\phi\dot{z},
$$
$$
w''=\psi''z+\psi'\phi\dot{z}+\psi'\phi\dot{z}+
\psi\phi'\dot{z}+\psi\phi^2\ddot{z}=
\psi''z+(2\psi'\phi+\psi\phi')\dot{z}+\psi\phi^2\ddot{z}. 
$$
Thus,
$$
w''-\l a^{-1}w=\psi\phi^2(\ddot{z}+\frac{2\psi'\phi+\psi\phi'}{\psi\phi^2}+
\frac{\psi''}{\psi\phi^2}z-\l \frac{a^{-1}\psi}{\psi\phi^2}z)=0.  
$$
Now we choose $\phi,\psi$ so that $2\psi'\phi+\psi\phi'=0$ and 
$\psi\phi^2=a^{-1}\psi$, which yields $\phi=a^{-1/2}$, $\psi:=a^{1/4}$. The 
equation then takes the form
$$
\ddot{z}+\frac{\psi''}{\psi\phi^2}z-\l z=
\ddot{z}+(a^{1/4})''a^{3/4}z-\l z=\ddot{z}-
[\frac3{16}a^{-1}(a')^2-\frac14a'']z-\l z=0.
$$
The boundary condition is
$$
w'(0)=(\frac14a^{-3/4}a'z+a^{-1/4}\dot{z})(0)= a^{-
1/4}(0)[\dot{z}(0)+\frac14a^{-1/2}(0)a'(0)]=0.
$$
Since these transformations are invertible, one concludes that $z$
satisfies 
(\ref{eqz}) if and only if $w$ satisfies (\ref{eqw}).
\end{proof}
\begin{lemma}\lb{comp} Consider two equations of the type (\ref{eqz}) with 
$q=q_i$ and $h=h_i$, where $i=1,2$. Then the set of products of the solutions 
$\{z_1(\cdot,\l)z_2(\cdot,\l)|\all\l>0\}$ is dense in $L^1([0,1])$.
\end{lemma}
\begin{proof} Let $z_0$  be a solution to (\ref{eqz}) with $q=0$. Then, up
to a 
multiplicative constant, one has: 
$z_0(\xi)=\cos(k\xi)+\frac{h}{k}\sin(k\xi)$, $\l=-k^2$. 
For $z_1$ and $z_2$ one has a representation through the transformation kernels 
$K_1,K_2$, which depend on $q_i$ and $h_i$:
$$
z_i(\xi,\l)=z_0(\xi,\l)+\Int^{\xi}_0K_i(\xi,\eta)z_0(\eta,\l)d\eta.
$$
The properties of the transformation operators can be found, e.g., in [1] 
and [5].
Also 
$$
z_i(\xi,\l)=\cos(k\xi)+h\Int^{\xi}_0\cos(ks)ds.
$$
Therefore, 
\begin{equation}\lb{zi}
z_i(\xi,\l)=\cos(k\xi)+\Int^{\xi}_0T_i(\xi,\eta)\cos(k\eta)d\eta,
\end{equation}
where
$$
T_i(\xi,\eta)=h_i+K_i(\xi,\eta)+h_i\Int^{\xi}_0K_i(\xi,s)ds.
$$
We have used above the following formula 
$$
\Int^{\xi}_0K_i(\xi,\eta) 
\Int^{\eta}_0\cos(ks)dsd\eta=\Int^{\xi}_0\cos(k\eta)(\Int^{\xi}_{\eta} 
K_i(\xi,s)ds)d\eta.
$$
Completeness of the set of products of the functions of the form (\ref{zi}) was 
proved in \cite{R4}.
\end{proof}
\begin{proof}[Proof of theorem \ref{propC}] Consider the set of products 
$w_1(\cdot,\l)w_2(\cdot,\l)$, where $w_i$ are the solutions to (\ref{eqw}) with 
$a=a_i$. By Lemma \ref{iso} 
$$
w_1(x,\l)w_2(x,\l)=(a_1(x)a_2(x))^{1/4}z_1(\xi(x),\l)z_2(\xi(x),\l).
$$
Define the linear operator $(Af)(x):=(a_1(x)a_2(x))^{1/4}f(\xi(x))$. This is a 
bounded linear operator from $L^1([0,\xi(1)]$ to $L^1([0,1])$, wchich has a 
bounded inverse since $a_i$ are continuous and positive and $\xi(x)$ is 
monotone. Also $A(z_1z_2)=w_1w_2$ and the same is true for linear combinations 
of the products. Since these combinations are dense in $L^1([0,\xi(1)]$ by
Lemma 
\ref{comp} and $A$ is a linear isomorphism we may conclude that linear 
combinations of products of $w_i$ are dense in $L^1([0,1])$. By Lemma 
(\ref{der}) $w_1w_2=v'_1v'_2$, where $v_i$ are solutions to \ref{eqvi} and the 
system of products of the derivatives is, therefore, complete.
\end{proof}

\section{Finding conductivity: uniqueness and non-uniqueness}\lb{s3}

\begin{theorem}\lb{uniq} There exists at most one solution to the inverse 
problem (\ref{ip1})-(\ref{ip4}).
\end{theorem}
\begin{proof} Define $v(x,\l):=\Int_0^\infty e^{-\l t}u(x,t)dt$. Applying the 
Laplace transform to equations (\ref{ip1})-(\ref{ip4}) one gets
\begin{equation}\lb{lip1}
(a(x)v')'-\l v=0,\ 0\leq x\leq 1,\ \l>0,
\end{equation}
\begin{equation}\lb{lip2}
v(0,\l)=0,\ v(1,\l)=F(\l):=\Int_0^\infty e^{-\l t}f(t)dt, 
\end{equation}
\begin{equation}\lb{lip3}
a(1)v'(1,\l)=G(\l):=\Int_0^\infty e^{-\l t}g(t)dt. 
\end{equation}
Assume that there exist $a_1$ and $a_2$ which generate the same data 
$f$ and $g$ 
and, therefore, the same $F$ and $G$. Put in equation (\ref{lip1}) $a=a_i$, 
$v=v_i$, $i=1,2$, and subtract the second equation from the first to get
\begin{equation}\lb{eqp}
(a_1(x)w')'-\l w=(p(x)v'_2)',
\end{equation}
where $p:=a_1-a_2$, $w:=v_1-v_2$. Let $\psi$ be an arbitrary solution to 
the problem:
\begin{equation}\lb{eqpsi}
(a_1(x)\psi')'-\l\psi=0,\ \psi(0,\l)=0.
\end{equation}
Multiply (\ref{eqp}) by $\psi$ and integrate by parts over $[0,1]$ to get:
$$
-\Int_0^1a_1w'\psi'dx+a_1w'\psi|^1_0-\l\Int_0^1w\psi dx=
-\Int_0^1pv'_2\psi'dx+pv'_2w|^1_0.
$$
Integrating the first term by parts once more and observing that the
non-integral terms vanish because
for $x=0$ and $x=1$ $w(x,\l)=0$ by (\ref{lip2}),
$\psi(0)=0$, and 
$a_1(1)v'_1(x,\l)=a_2v'_2(1,\l)$ by (3.3), 
one obtains an orthogonality relation:
$$
0=\Int_0^1pv'_2\psi'dx \quad \forall\l>0.
$$
 By Theorem \ref{propC} the set of products 
$\{v'_2(\cdot,\l)\psi'(\cdot,\l)\}$ is complete in $L^1(0,1)$. Therefore, 
$p=0$ and $a_1=a_2$.
\end{proof}
As we have mentioned in the Introduction, the uniqueness of the solution
to the 
inverse problem fails to hold if the flux measurements are taken at the
left end of 
the interval.
\begin{theorem}\lb{noniq} Consider problem (\ref{ip1})-(\ref{ip3}) with  
extra data (\ref{ip5}). Then the conductivities $a(x)$ and $a(1-x)$ produce the 
same flux $h(t)$.
\end{theorem}
\begin{proof} Laplace transform turns problem (\ref{ip1})-(\ref{ip3}) into 
(\ref{lip1})-(\ref{lip2}) with the extra data
\begin{equation}\lb{lip5}
a(0)v'(0,\l)=H(\l):=\Int_0^\infty e^{-\l t}h(t)dt. 
\end{equation}
Make the substitution $b:=av'$ and write the problem in the 
form
$$
b''-\l a^{-1}(x)b=0,
$$
$$
b'(0,\l)=0,\ \ \b'(1,\l)=\l F(\l),\ \ b(0,\l)=H(\l).
$$
Introduce $\theta(x,\l)$ as the solution to the problem
\begin{equation}\lb{eqth}
\theta''-\l a^{-1}(x)\theta b=0,\ \ \theta'(0,\l)=0,\ \ \theta(0,\l)=1. 
\end{equation}
Clearly, $b(x,\l)=c(\l)\theta(0,\l)$ and, on the one hand,
$$
H(\l)=b(0,\l)=c(\l)\theta(0,\l)=c(\l),
$$
while on the other hand,
$$
\l F(\l)=b'(1,\l)=c(\l)\theta'(1,\l)=H(\l)\theta'(1,\l).
$$
Thus, 
\begin{equation}\lb{H}
H(\l)=\frac{\l F(\l)}{\theta'(1,\l)}.
\end{equation}
The function $\theta'(1,\l)$ is an entire function of exponential type, which is 
determined by its zeros $\l_j$ by the Hadamard formula. The relation 
$\theta'(1,\l_j)=0$ together with (\ref{eqth}) imply that $\theta(x,\l_j)$
are the
eigenfunctions of the Neumann problem for
equation  (\ref{eqth}) and $\l_j$ are the 
corresponding eigenvalues. If some $\phi(x)$ satisfies
differential equation  (\ref{eqth}) and the 
Neumann boundary 
conditions then $\phi(1-x)$ satisfies
equation (\ref{eqth}) with $a(x)$ replaced by $a(1-
x)$ and the Neumann boundary conditions. In other words, the set
$\{\l_j\}$ will be 
the same for $a(x)$ and $a(1-x)$.
Since the set $\{\l_j\}$ determines $\theta'(1,\l)$ uniquely, this function for 
$a(x)$ 
and $a(1-x)$ is the same. But then by (\ref{H}) and (\ref{lip5}) the functions 
$h(t)$ are also the same.
\end{proof}
Thus, if measurements are taken on a wrong end of the interval, the unique 
recovery of the conductivity is impossible, in general. If $a(x)\neq a(1-x)$ 
then these two different functions produce the same flux data.
\begin{remark} The situation is similar to the one for the Sturm-Liouville 
equation:
\begin{equation}\lb{StL}
v''-q(x)v+\l v=0,\ v(0,\l)=0,\ v(1,\l)=F(\l),
\end{equation}
and the extra data $v'(0,\l)=G(\l)$. In this case $q(x)$ and $q(1-x)$ will also 
produce the same extra data $G$. Even more can be said in this case: $q$ is 
determined by $G$ on $[0,1/2)$. This follows from the old result in \cite{HoLi}: 
$q$ on the half of the interval can be recovered from the knowledge of one 
spectrum 
for problem (\ref{StL}) with Dirichlet boundary conditions. It also 
follows from Theorem 3.1 in \cite{R3}.
\end{remark}



\begin{thebibliography}{99}

\bibitem{Lev1}B.M. Levitan,{\it Generalized translation operators and some of 
their applications}, Jerusalem, Israel Program for Scientific Translations, 
1964.


\bibitem{HoLi}H.Hochshtadt, B.Lieberman, {\it An inverse Sturm-Liouville problem 
with mixed data}, SIAM J.Appl.Math, 34 (1978), 331-348.

\bibitem{R3}A.G.Ramm, {\it  Property C for ODE and applications to inverse 
problems}, in the book {\it Operator Theory and Its Applications},
Amer. Math. Soc., Fields Institute Communications vol. 25 (2000), pp.15-75, 
Providence, RI.

\bibitem{R4}A.G.Ramm, {\it Property C for ODE and applications to inverse
scattering}, Zeit. fuer Angew. Analysis, 18, no.2 (1999), 331-348.

\bibitem{R5}A.G.Ramm, {\it  Multidimensional inverse scattering
problems}, Longman/Wiley, New York, 1992, 1-385.

\bibitem{R6}A.G.Ramm, {\it Finding conductivity from boundary
measurements}, Comp.\& Math.with Appl., 21, no.8 (1991), 85-91.

\bibitem{R7}A.G.Ramm, {\it An inverse problem for the heat equation} 
Jour. of Math. Anal. Appl., 264, N 2, (2001), 691-697.

\end{thebibliography}
\end{document}